\documentclass[prl,showpacs,amsmath,assymb,eufrak]{revtex4}
\usepackage{bm}
\usepackage{epsfig,amsfonts,amsbsy}
\newcommand{\sbkt}[1]{\langle#1\rangle}
\newcommand{\bbkt}[1]{\bigl\langle#1\bigr\rangle}

\newcommand{\thetae}{\theta_\mathrm{ex}}
\newcommand{\thetast}{\theta_\mathrm{st}}
\newcommand{\Thetae}{\Theta_\mathrm{ex}}

\newcommand{\Phie}{\Phi_\mathrm{ex}}
\newcommand{\Xe}{X_\mathrm{ex}}

\newcommand{\Jst}{J^\mathrm{st}}

\newcommand{\Di}{\mathit{\Delta}}

\newcommand{\Tbar}{\bar T}
\newcommand{\TL}{(1+\frac{\varepsilon}{2})\Tbar}
\newcommand{\TR}{(1-\frac{\varepsilon}{2})\Tbar}
\newcommand{\Tb}{T^\mathrm{b}}

\newcommand{\Sleq}{ S_\mathrm{leq}}
\newcommand{\Seq}{ S_\mathrm{eq}}
\newcommand{\Sneq}{ S_\mathrm{neq}}
\newcommand{\dTheta}{\Theta^{\dagger}}
\begin{document}
\title{Numerical examination of steady-state thermodynamics from the entropy connected to the excess heat}

\author{Yoshiyuki Chiba and Naoko Nakagawa}
\date{\today}

\begin{abstract}
We numerically determine the entropy for heat-conducting states,
which is connected to the so-called excess heat considered as a basic quantity for steady-state thermodynamics in nonequilibrium.
We adopt an efficient method to estimate the entropy from the bare heat current
and find that the obtained entropy agrees with the familiar local equilibrium hypothesis well.
Our method possesses a wider applicability than local equilibrium and
opens a possibility to compare thermodynamic properties of complex systems with those in the local equilibrium. 
We further investigate the entropy for heat-conducting states and find that it exhibits both extensive and additive properties;
however, the two properties do not degenerate each other differently from those at equilibrium.
The separation of the extensivity and additivity makes it difficult to apply
 powerful thermodynamic methods.
\end{abstract}


\maketitle

\section{Introduction}
The extension of  thermodynamics from equilibrium to nonequilibrium states has been a goal for a long time, and
 several attempts to extend equilibrium thermodynamics to nonequilibrium steady states (NESSs) have been made.
One of these attempts is the extension of equilibrium thermodynamic functions 
to the NESSs.
For this purpose, the extension of thermodynamic relations 
was proposed to transition between two NESSs
\cite{Landauer, Oono-Paniconi, Ruelle, Sasa-Tasaki,Hatano-Sasa,KNST,KNST-nl,Saito-Tasaki,NN2012,KNST-exact,BGJLL2013,MaesNetocny2014}.
Landauer tried to formulate thermodynamics far from equilibrium by taking part of the heat accompanied with the change in states \cite{Landauer}.
In a similar sprit, Oono and Paniconi proposed steady-state thermodynamics (SST) by adopting the ``{\it excess heat}" instead of the bare heat \cite{Oono-Paniconi},
in which the excess heat is proposed as the heat after subtracting the housekeeping heat for maintaining the steady state.

Sasa and Tasaki have constructed their SST from a phenomenological point of view \cite{Sasa-Tasaki}.
Taking heat-conducting or shared systems as examples,
they argued that the local steady states differ from the local equilibrium states for the system  in the thermodynamic limit.
Assuming the extensivity of the thermodynamic functions as is in equilibrium,
they proposed their framework of SST.
The proposed SST is rather powerful, as various equilibrium properties are kept, 
such as the convexity of the thermodynamic functions or the second law.
By utilizing these aspects, they predicted what they call flux-induced osmosis, 
which has not been experimentally examined.

From a microscopic point of view, 
Hatano and Sasa have derived a Clausius-like inequality 
with their definition of the excess heat
\cite{Hatano-Sasa}.
The Shannon entropy difference is connected to their excess heat in the quasistatic limit.
Their inequality is clear and mathematically rigorous;  however,
their excess heat is rather difficult to access experimentally.
Moreover, their decomposition of the heat into the excess and housekeeping parts is applicable to a system described by even variables 
but not straightforwardly applied to a system with odd variables \cite{Spinney-Ford,Spinney-Ford2, Sasa2013}.
That is, the two inequalities for the excess and housekeeping heats seem to be valid when the system's probability density is time-symmetric,
whereas even the definition of the excess heat becomes unclear in a system with degrees of freedom for the momentum.

Komatsu et al. have derived another form of the Clausius-like equality that holds in the quasistatic limit
by defining the excess heat as the one after subtracting the expected steady heat currents at each moment \cite{KNST}.
Their equality holds regardless of the time symmetry of the probability density,
and the entropy extended to the NESSs is connected to the symmetric Shannon entropy, 
which becomes different from the usual Shannon entropy in a system with odd variables.
Their excess heat is experimentally observable and connected to an entropy in their Clausius equality. 
When the probability density becomes time-symmetric, their excess heat can estimate the Shannon entropy 
so that the equality has a correspondence with the quasistatic limit of the Hatano-Sasa equality.
On the other hand, since the definition of the excess heat is slightly different from the one of  Hatano and Sasa, 
the Clausius equality does not always have a corresponding inequality.

We here raise questions about the proposed thermodynamic functions in these SST: 
How is the entropy  extended to the NESSs,
or what kind of properties are kept or altered in the change from equilibrium to nonequilibirum?
Is it related to the local equilibrium or the local steady states that Sasa and Tasaki assumed?
To examine such questions, we need to obtain experimental observations apart from the theoretical issues.
We thus perform numerical experiments for the NESSs in this study.
Noting experimental observability,
we focus on  the excess heat adopted by Komatsu et al. 
and determine the entropy in the NESSs according to their Clausius relation.
We then explore the basic properties of the entropy in the NESSs 
and examine the validity of the assumptions by Sasa and Tasaki. 
We compare the numerically determined entropy with the one determined according to the local equilibrium hypothesis in order to examine the possibility of local steady states.

Following this introduction, we summarize our numerical results for one-dimensional heat-conducting systems.
First, we introduce an effective method to determine the entropy connected to the excess heat.
Comparing our numerically determined entropy with the entropy deduced from the local equilibrium hypothesis,
we find a good correspondence between the two entropies.
Thus, we do not observe a difference between the local equilibrium and the local steady states.
The correspondence is confirmed for systems in the thermodynamic limit and for those of rather small sizes. 
Even though we cannot solve the steady states for the systems of small sizes due to the complexity of the finite size effects, the entropy measured at the boundary of the system traces the steady states appropriately.

Next, we examined the extensivity and additivity of the obtained entropy
and found that  both the extensivity and additivity were maintained; 
however, the obtained extensivity 
was different from the one assumed by Sasa and Tasaki in \cite{Sasa-Tasaki}.
The extensivity and additivity are not unified in the NESSs, 
whereas the two properties are degenerated in equilibrium extensive systems.
Remembering that this degeneracy brings several powerful tools to the  equilibrium thermodynamics,
the SST based on the excess heat may not be so powerful as the equilibrium thermodynamics even though  it reproduce the thermodynamics of the local equilibrium hypothesis.

\section{Entropy connected to the excess heat}

\begin{figure}[t]
\centering
\begin{picture}(300,100)
\includegraphics[scale=0.4]{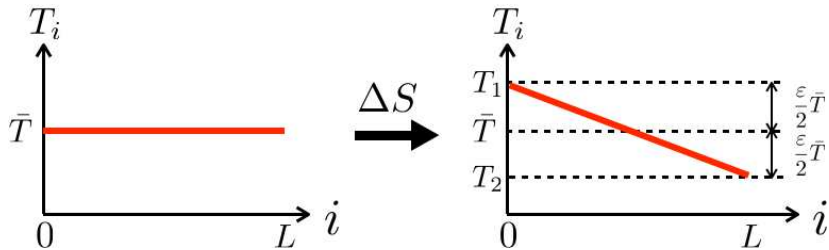}
%\begin{picture}(250,100)
%\includegraphics[scale=0.31]{Protocol.eps}
\end{picture}
\caption{Sketch of the temperature profile before the change (initial equilibrium state) and
in the final heat-conducting state. 
To determine the entropy in the heat-conducting state (right), 
 we apply a protocol to change the temperature of the heat baths from $(\Tbar, \Tbar)$ to $(T_1, T_2)$. We fix $\Tbar=\frac{T_1-T_2}{2}$.
}
\label{fig:protocol}
\end{figure}

We here explain the extended Clausius equality proposed in \cite{KNST}
by taking a one-dimensional heat-conducting lattice as an example.
The two ends of the lattice
 are in contact with heat baths at temperatures of $\Tb_1(t)$ and $\Tb_2(t)$, which can be dependent on the time $t$.
We measure the heat current $j_k(t)$ from the $k$th heat bath to the lattice.
When we fix the two temperatures as $T_1$ and $T_2$ and relax the system for a sufficiently long time, 
the system reaches a unique steady state with the steady heat current 
\begin{align}
\Jst(T_1, T_2, L):=\langle j_1\rangle=-\langle j_2\rangle,
\end{align}
where $\langle \cdot \rangle$ is the ensemble and/or long-time average
in the steady state.

In the following, we treat the change in the entropy 
from an equilibrium state at $\Tbar$ to a heat-conducting state with $T_1$ and $T_2$, where we impose $\Tbar=\frac{T_1+T_2}{2}$.
See Fig.~\ref{fig:protocol}.
We set  a nondimensional parameter
\begin{align}
\varepsilon = \frac{T_1-T_2}{\Tbar}
,
\end{align}
which is used to indicate the degree of nonequilibrium in the final heat-conducting state.
Note that $T_1=\TL$ and $T_2=\TR$.

We fix the protocol in the period $0\le t\le\tau$ to change the temperatures of two heat baths.
We assume that the lattice system is in equilibrium at $t=0$ and in the steady heat-conducting state  at $t=\tau$, where $\tau$ is much longer than the system's relaxation time.
Repeating the same experiments in the fixed protocol and measuring the heat current at each moment,
we obtain the temporal probability density for each time $t$.
Taking the ensemble average for each time $t$, 
the entropy production rates in the heat baths are defined as
\begin{eqnarray}
\theta(t):=-\frac{\bbkt{j_1(t)}}{\Tb_1(t)}-\frac{\bbkt{j_2(t)}}{\Tb_2(t)}
.
\end{eqnarray}
$\theta(t)$ can be decomposed into the expected steady contribution and the rest,
which is called the excess entropy production rate
\begin{eqnarray}
\thetast(t)&:=&-{\Jst(\Tb_1(t), \Tb_2(t), L)}\left(\frac{1}{\Tb_1(t)}-\frac{1}{\Tb_2(t)}\right),
\label{e:ep-st}\\
\thetae(t)&:=&\theta(t)-\thetast(t).
\label{e:ep-ex}
\end{eqnarray}
Consistent with the second law of thermodynamics, 
the steady part $\thetast$  is always nonnegative, 
whereas the excess part $\thetae$ is not necessarily bounded to nonnegative values.
Note that $\thetae$ vanishes in the steady state and is produced by an external operation.

The change in the entropy from  isothermal to  heat-conducting states 
is connected to the excess entropy production via the extended Clausius equality derived in \cite{KNST}.
It is given by
\begin{eqnarray}
\Sneq(L,\Tbar, \varepsilon)-\Seq(L,\Tbar) + \Thetae = O(\varepsilon^3),
\label{e:S}
\end{eqnarray}
which is valid in quasistatic operations.
Here, $\Thetae$ is the excess entropy production in the two heat baths:
\begin{eqnarray}
\Thetae=\int_0^{\tau} \thetae(t) ~dt ,
\end{eqnarray}
for the observation in the period $[0, \tau]$.

\section{Alternative equality effective for the experimental measurement of  $\Di\Sneq$}

\begin{figure}
\centering
\begin{picture}(300,100)
\includegraphics[scale=0.43]{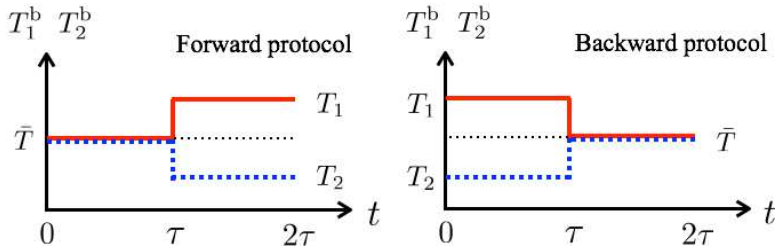}
%\begin{picture}(240,90)
%\includegraphics[scale=0.33]{FwdBwd.eps}
\end{picture}
\caption{
Pair of protocols for estimating $\Di\Sneq$ by utilizing \eqref{e:S2}.
The heat baths $\Tb_1$ and $\Tb_2$ are changed in a stepwise manner,
 as indicated by the red solid and blue dotted lines, respectively.
The forward and  backward protocols are exactly time-reversed.
The period $\tau$ is set to be much longer than the typical relaxation time.}
\label{fig:FwdBwd}
\end{figure}

The equality in \eqref{e:S} offers an experimental method to determine the entropy in the NESSs  by measuring the heat currents.
In the experiments, we have two difficulties: 
To obtain $\Sneq(L,\Tbar,\varepsilon)$, we need to measure $\Thetae$ in a pre-fixed quasistatic protocol, i.e., measure both $\theta(t)$ and $\thetast(t)$ at each moment 
(see the definition in \eqref{e:ep-ex}).
Honestly, such measurements are too difficult to complete.
In addition to the difficulties associated with performing the quasistatic operation,
there are also difficulties associated with performing a huge number of experiments 
to obtain $\thetast(t)$ for each moment $t$.
We should perform a series of experiments for each steady state driven by  $\Tb_1(t)$ and $\Tb_2(t)$.
After completing these measurements,
we at last reach $\Thetae$ to determine the entropy difference $\Di\Sneq := \Sneq-\Seq$.

To reduce such experimental difficulties, one may avoid the application of  quasistatic protocols and
adopt protocols consisting of a few stepwise changes, as shown on the left-hand side of Fig.~\ref{fig:FwdBwd},
because the number of the internal steady states is suppressed to a few.
The problem with these protocols is how to estimate the deviation from the quasistatic limit.

Considering such difficulties, we adopt another version of the extended Clausius equality
\begin{eqnarray}
\Sneq(L,\Tbar, \varepsilon)-\Seq(L,\Tbar) 
=
-\frac{1}{2}(\Theta -\dTheta)+ O(\varepsilon^3)
,
\label{e:S2}
\end{eqnarray}
from which the equality \eqref{e:S} is obtained in the quasistatic limit
as is explained in the Appendix or in \cite{KNST}.
The advantage of \eqref{e:S2} is its applicability to ``finite speed" protocols. 
This equality is not for a single quasistatic protocol but for a pair of protocols
corresponding to the forward and backward protocols  exemplified in Fig.~\ref{fig:FwdBwd}.
In each protocol, we determine the entropy production in the heat baths,
$\Theta=\int_0^{\tau} \theta(t) dt$ and $\dTheta=\int_0^{\tau}\theta^\dagger(t) dt$.
$\theta(t)$ and $\theta^\dagger(t)$ are the entropy production rates at time $t$ along the forward and backward protocols, respectively.

We here introduce $\delta$, which indicates the magnitude of the ``{\it deviation from the quasistatic limit,}"
\begin{align}
\delta := \frac{\Thetae+\Thetae^\dagger}{\varepsilon}
.
\label{e:delta}
\end{align}
Note that $\delta=0$ in the quasistatic limit,
whereas $\Thetae^\dagger=-\Thetae+\varepsilon\delta$ in general protocols.
In protocols that change from equilibrium to the heat-conducting state,
we have $\delta=O(\varepsilon)$, as shown in the Appendix.
Remembering that  $\theta_\mathrm{st}(t)$ is common in the pair protocols, we have $\Theta-\dTheta=\Thetae-\Thetae^\dagger$.
Then, the equality in \eqref{e:S2} is transformed as
\begin{align}
\Sneq(L,\Tbar, \varepsilon)-\Seq(L,\Tbar) 
+\Thetae
=
O(\varepsilon\delta, \varepsilon^3)
.
\label{e:S3}
\end{align}
Equation \eqref{e:S3} reduces to \eqref{e:S} by limiting the operation to the quasistatic limit.
Comparing \eqref{e:S2} with \eqref{e:S3}, 
the equality in \eqref{e:S2} for  finite-speed operations
offers the same precision for determining $\Sneq$
with the equality in \eqref{e:S} for quasistatic operations.

The equality in \eqref{e:S2} has another advantage, i.e.,
it does not require the measurement of $\theta_\mathrm{st}(t)$, as was pointed out in \cite{KNST-numerical}.
In smooth protocols, \eqref{e:S}  requires a huge number of experiments to determine $\Thetae$;
however, \eqref{e:S2} only requires  two experiments to form a pair.

\section{Numerical experiment}

\subsection{Model}

%model
As a model system for performing numerical experiments, we adopt a one-dimensional heat-conducting lattice.
We shall take the $\phi^4$ model, which is known to exhibit normal heat conduction obeying Fourier's law \cite{Aoki-Kusnezov}.
Its Hamiltonian is written as
\begin{equation}
H(\{x\},\{p\})=\sum_{i=1}^L \left(\frac{p_i^2}{2} +\frac{x_i^4}{4}\right)+\sum_{i=1}^{L-1}\frac{(x_{i+1}-x_i)^2}{2},
\end{equation}
where $x_i$ and $p_i$ are the deviation from the equilibrium position and the momentum for the $i$th site, respectively; and
$L$ is the number of elements in the lattice.
The two ends, $i=1$ and $L$,  of the lattice
 are in contact with heat baths at the temperatures of $\Tb_1(t)$ and $\Tb_2(t)$, respectively.
The evolution equation of each element
is given by the canonical equations for the bulk ($2\leq i \leq L-1$)
and the Langevin equations for the two ends:
\begin{eqnarray}
\dot p_1 &=& -\frac{\partial H}{\partial x_1} -\gamma p_1 +\sqrt{2\gamma \Tb_1(t)}\circ \xi_1(t), \quad \dot x_1=p_1,
\label{e:Langevin1}\\
\dot p_i &=& -\frac{\partial H}{\partial x_i}, \quad \dot x_i=p_i,
\label{e:canonical}\\
\dot p_L &=& -\frac{\partial H}{\partial x_L} -\gamma p_L +\sqrt{2\gamma \Tb_2(t)}\circ \xi_2(t), \quad  \dot x_L=p_L,
\label{e:LangevinL}
\end{eqnarray}
where $\circ$ is Stratonovich's multiplier and $k_\mathrm{B}$ is taken as unity.
$\xi_k(t)$ is Gaussian white noise satisfying $\langle \xi_k(t)\rangle=0$
and $\langle \xi_k(t)\xi_{k'}(t')\rangle=\delta_{k k'}\delta(t-t')$.
We set $\gamma=1$.

\subsection{Measurement of the entropy productions}

In order to apply the relation in \eqref{e:S2} to determine $\Sneq$, 
we need to measure $\Theta$ and $\dTheta$ for the forward and  backward protocols, respectively.
For this, we should observe the heat current $j_k(t)$ from the $k$th heat bath to the lattice.
In the present model with a Langevin thermostat, the heat currents are defined as
\begin{eqnarray}
j_1(t) &:=& p_1 \circ\left(-\gamma p_1 +\sqrt{2\gamma \Tb_1(t)} \circ\xi_1(t)\right),\\
j_2(t) &:=& p_L \circ\left(-\gamma p_L +\sqrt{2\gamma \Tb_2(t)} \circ\xi_2(t)\right)
,
\label{e:j1j2}
\end{eqnarray}
adopting stochastic energetics \cite{Sekimoto-JPSJ,Sekimoto}.
These heat currents are the only observable in the following numerical measurement.

We prepare the equilibrium ensemble at $\Tbar$ to start the forward protocol in Fig.~\ref{fig:FwdBwd}.
Taking this as the initial ensemble  at $t=0$, 
we integrate the evolution equations in \eqref{e:Langevin1}, \eqref{e:canonical}, and \eqref{e:LangevinL}.
We abruptly change the temperature of the two heat baths at $t=\tau$
and continue the integration up to $2\tau$,
where $\tau$ is much longer than the system's relaxation time.
We measure $j_1(t)$ and $j_2(t)$ along each trajectory
and take their ensemble averages to form 
\begin{align}
\Theta=-\int_{\tau}^{2\tau} dt~ \left(\frac{\bbkt{j_1(t)}}{T_1}+\frac{\bbkt{j_2(t)}}{T_2}\right).
\end{align}
 The contribution to $\Theta$ in the period $[0, \tau]$ should vanish because it is in equilibrium.

The procedure is almost the same in the backward protocol,
except for the initial ensemble, which should be prepared in the steady heat-conducting states.
We prepare the steady heat-conducting states of $T_1$ and $T_2$
and then change  both temperatures to $\Tbar$ at $t=\tau$.
We measure the entropy production 
\begin{align}
\dTheta=-\int_{0}^{\tau} dt ~\left(\frac{\bbkt{j_1(t)}^\dagger}{T_1}+\frac{\bbkt{j_2(t)}^\dagger}{T_2}\right) 
-\int_{\tau}^{2\tau} dt ~\frac{\bbkt{j_1(t)}^\dagger +\bbkt{j_2(t)}^\dagger}{\Tbar }
,
\end{align}
where $\bbkt{\cdot}^\dagger$ means the ensemble average along the backward protocol.
The first integral gives the steady entropy production in the heat-conducting state,
and the second corresponds to the excess entropy production induced by the change from heat-conducting  to equilibrium states.
Therefore, the two excess entropy productions are written as
\begin{align}
\Thetae&=-\int_{\tau}^{2\tau} dt~ \left(\frac{\bbkt{j_1(t)}}{T_1}+\frac{\bbkt{j_2(t)}}{T_2}\right)+\int_{0}^{\tau} dt ~\left(\frac{\bbkt{j_1(t)}^\dagger}{T_1}+\frac{\bbkt{j_2(t)}^\dagger}{T_2}\right), \\
\Thetae^\dagger &=-\int_{\tau}^{2\tau} dt ~\frac{\bbkt{j_1(t)}^\dagger +\bbkt{j_2(t)}^\dagger}{\Tbar}
.
\end{align}
According to the relation in \eqref{e:S3} for the forward and backward protocols,
the two excess entropy productions are respectively connected to the entropy difference as
\begin{align}
\Sneq(L,\Tbar, \varepsilon)-\Seq(L,\Tbar) &= -\Thetae + O(\varepsilon\delta,\varepsilon^3),\\
\Sneq(L,\Tbar, \varepsilon)-\Seq(L,\Tbar) &=\Thetae^\dagger+ O(\varepsilon\delta,\varepsilon^3),
\end{align}
Figure \ref{fig:EP} plots $-\Thetae$ and $\Thetae^\dagger$ 
simultaneously for various values of $\varepsilon$.
The equilibrium state is common
for each value of $\varepsilon$, but the heat-conducting states are different.
It is remarkable that $\Thetae^\dagger$ greatly deviates from $-\Thetae$ for each value of $\varepsilon$,
 whereas these two should coincide with each other in the quasistatic limit.
Such deviation indicates that $\Thetae$ cannot give
information about $\Di\Sneq$.

\subsection{Entropy in the local equilibrium hypothesis}

\begin{figure}
\centering
\begin{picture}(220,150)
%\put(-10,100){(a)}
\includegraphics[scale=0.8]{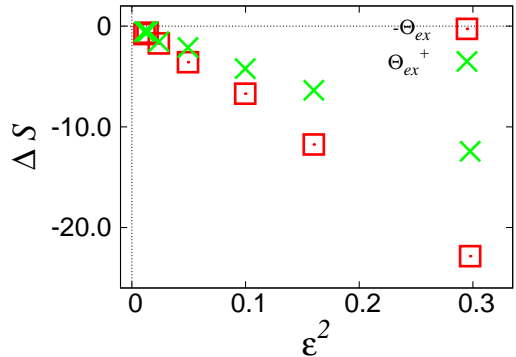}
\end{picture}
\caption{The dependence of the degree of nonequilibiurm $\varepsilon$ for the excess entropy productions $-\Thetae$ and $\Thetae^\dagger$ for $L=500$.
Each error bar is smaller than the size of each point.
$(T_1,T_2)=$
$(0.5,0.45)$, $(0.45,0.4)$, $(0.7,0.6)$, $(0.5,0.4)$, $(0.55,0.4)$, $(0.6,0.4)$, $(0.7,0.4)$.
$-\Thetae$ and $\Thetae^\dagger$ should coincide in the quasistatic limit; however, the deviation becomes significant in the one-step protocol, especially for larger $\varepsilon$.
}
\label{fig:EP}
\end{figure}

In the previous section, we found that  $\Thetae$ did not 
predict $\Di\Sneq$ when we perform  the one-step protocol in Fig.~\ref{fig:FwdBwd}.
We now focus on $-(\Theta-\dTheta)/2$ as a candidate of $\Di\Sneq$, 
which corresponds to the middle point of $-\Thetae$ and $\Thetae^\dagger$ in Fig.~\ref{fig:EP}.

From the viewpoint of examining the proposed SST, we compare the obtained $\Di\Sneq$ with the entropy difference deduced from the local equilibrium hypothesis.
In particular, we are interested in whether the local steady states proposed by Sasa and Tasaki are detected differently from the local equilibrium states according to the proposed SST.
As is well known, 
the hypothesis of local equilibrium has been successful in several nonequilibrium systems
in the thermodynamic limit.
%Lebowitz and Spohn have shown that the nonequilibriurm states near equilibrium
%are described by the local equilibrium quantities 
%for a system obeying the Galavotti-Cohen fluctuation theorem \cite{Lebowitz-Spohn}.
Examination of  the relation  of $\Sneq$ to the local equilibrium hypothesis
 could be a good test for the proposed SST.
It may offer a check of the validity of the method of measuring $\Sneq$.
Thus, the purposes of the present comparison are as follows:
\begin{description}
\item{(i)} examine whether the SST developed in the spirit of the excess heat agree with the local equilibrium hypothesis.
\item{(ii)}  check the precision of $-(\Theta-\dTheta)/2$ for the estimate of $\Sneq$.
\end{description}

As a starting point for introducing the local equilibrium hypothesis, 
we assume that the local temperature corresponds to the kinetic temperature
for each $i$th site as
\begin{eqnarray}
T_i=\langle p_i^2\rangle.
\label{e:Tleq}
\end{eqnarray}
In our numerical simulation, the specific heat $c$ is obtained to be approximately constant, 
and $c\simeq 0.88$ for a wide range of temperatures, 
which is consistent with the reported value of $0.86$ for a sufficiently long lattice \cite{Aoki-Kusnezov}.
Then, 
the local internal energy and local entropy are expressed as
\begin{align}
u_i=c T_i, \qquad s_i = c \log T_i
,
\label{e:leq}
\end{align}
where 
the formula for the local entropy is derived from $d s_i =  d u_i / T_i$.
Then, we define the entropy for the entire system as
\begin{align}
S_\mathrm{leq} (L,\Tbar,\varepsilon)&=\sum_{i=1}^L s_i
.
\label{e:Sleq}
\end{align}

To obtain $\Di\Sleq$, we calculate both $\Sleq(L,\Tbar,\varepsilon)$ in the steady heat-conducting state  and $\Seq(L,\Tbar)$ in the equilibrium state.
In each steady state, 
we estimate the local kinetic temperature $T_i$ by a long-time and/or ensemble average and then calculate $s_i$ by applying \eqref{e:leq}. 

\begin{figure}
\centering
\begin{picture}(220,150)
%\put(-10,100){(a)}
\includegraphics[scale=0.8]{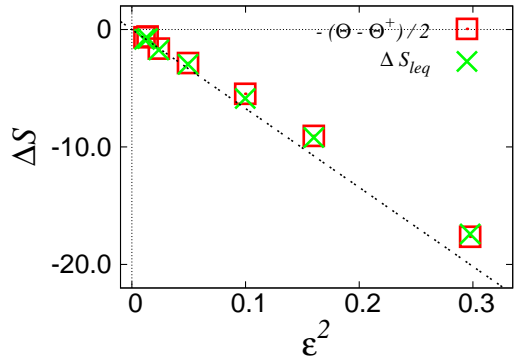}
\end{picture}
\caption{The dependence of the degree of nonequilibiurm $\varepsilon$
for the entropy difference $-\frac{1}{2}(\Theta-\dTheta)$ and $\Di\Sleq$ for $L=500$.
Each error bar is smaller than the size of each point.
$(T_1,T_2)=$
$(0.5,0.45)$, $(0.45,0.4)$, $(0.7,0.6)$, $(0.5,0.4)$, $(0.55,0.4)$, $(0.6,0.4)$, $(0.7,0.4)$.
The dotted line corresponds to the thermodynamic limit in \eqref{e:SleqInf}.
The coincidence of $\Di\Sleq\simeq -\frac{1}{2}(\Theta-\dTheta)$ has been maintained over the range of $\varepsilon$ in spite of the large deviation between $\Thetae$ and $\Thetae^\dagger$ in Fig.~\ref{fig:EP}.
}
\label{fig:SleqSneq}
\end{figure}

\begin{figure}[t]
\centering
\begin{picture}(220,150)
%\put(-10,100){(a)}
\includegraphics[scale=0.8]{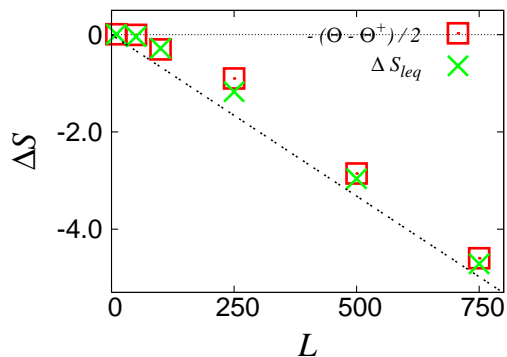}
\end{picture}
\caption{System-size dependence of the entropy difference $-\frac{1}{2}(\Theta-\Theta^\dagger)$ and $\Di\Sleq$
for the change $(\Tbar,\Tbar)$ to $(T_1, T_2)$.
$(T_1, T_2)$ is fixed at $(0.5, 0.4)$, whereas
$L=10$, $50$, $100$, $250$, $500$, $750$.
Each error bar is smaller than the size of each point.
The dotted line corresponds to the thermodynamic limit in \eqref{e:SleqInf}.
Note that $\Di\Sneq$ agrees with $\Di\Sleq$, even for the smallest $L$.
}
\label{fig:extensivity}
\end{figure}

\subsection{Numerical results for $\Sneq$}

Figure \ref{fig:SleqSneq} plots the numerical results for
$-(\Theta-\dTheta)/2$ in \eqref{e:S2} 
and 
$\Di \Sleq:=\Sleq(L,\Tbar,\varepsilon)-\Seq(L,\Tbar)$
simultaneously
for various values of $\varepsilon$ for the lattice of $L=500$.
In Fig.~\ref{fig:extensivity},
we plot the same quantities for various values of $L$ and a fixed $\varepsilon$ corresponding to $T_1=0.5$ and $T_2=0.4$,
i.e, $\Tbar=0.45$ and $\varepsilon=0.22$. 
We note in both Figs.~\ref{fig:SleqSneq} and \ref{fig:extensivity}
that the two quantities coincide well with each other for all values of $\varepsilon$ and $L$,
i.e.,
\begin{align}
\Sleq(L,\Tbar,\varepsilon)-\Seq(L,\Tbar) \simeq -\frac{1}{2}(\Theta-\dTheta)
.
\end{align}
This coincidence gives a plausible conclusion  for the examination of (i) and (ii) as
\begin{align}
{\rm (i)~} \qquad& \Sneq(L,\Tbar,\varepsilon)\simeq\Sleq(L,\Tbar,\varepsilon),\\
{\rm(ii)}\qquad& \Sneq(L,\Tbar,\varepsilon)-\Seq(L,\Tbar) \simeq -\frac{1}{2}(\Theta-\dTheta).
\end{align}
Thus, we have the conclusion 
that $-\frac{1}{2}(\Theta-\dTheta)$ consistently gives as good an estimate for $\Sneq$ 
as consistently with the  prediction in \eqref{e:S2},
even if the applied protocol is a stepwise change far from the quasistatic limit.
Remembering the large deviation of $-\Thetae$ and $\Thetae^\dagger$ in Fig.~\ref{fig:EP},
the present coincidence in Fig.~\ref{fig:SleqSneq}  is surprisingly good.
We also emphasize that Fig.~\ref{fig:SleqSneq} contains a point
where $|\varepsilon|\simeq 0.55$ with $T_1=0.7$ and $T_2=0.4$,
which is not very small, but $-\frac{1}{2}(\Theta-\dTheta)$ still traces $\Di\Sleq$ well.
Thus, we conclude that $\Di\Sneq$ in the NESSs
can be experimentally predicted by utilizing \eqref{e:S2},
 and moreover, that the local quantities connected to the local entropy can become accessible 
by the relation in \eqref{e:S2} by the measurement of the heat current.

As we have succeeded in obtaining $\Sneq$ numerically, we now proceed to examine its properties.
Below, we describe $-\frac{1}{2}(\Theta-\dTheta)$ as $\Di\Sneq$ according to \eqref{e:S2}.
Figure \ref{fig:SleqSneq} shows 
$\Di \Sneq$  for various values of $T_1$ and $T_2$ with $L=500$.
Both $\varepsilon$ and $\Tbar$ vary from point to point,
but $\Di\Sneq$ shows a clear dependence on only $\varepsilon$.
This means that $\Di\Sneq$ is proportional to $\varepsilon^2$ and independent of  $\Tbar$. 
In Fig.~\ref{fig:extensivity},
We observe that $\Di\Sneq$ is proportional to $L$ for large $L$,
in which the system is closer to equilibrium with a lower temperature gradient.
Summarizing, we conclude that
\begin{align}
\Sneq(L,\Tbar,\varepsilon) \simeq\Seq(L,\Tbar) - a L \varepsilon^2
\label{e:scale}
\end{align}
for larger values of $L$ where $a$ is a certain constant.

\begin{figure}[t]
\centering
\begin{picture}(220,150)
%\put(-10,100){(a)}
\includegraphics[scale=0.8]{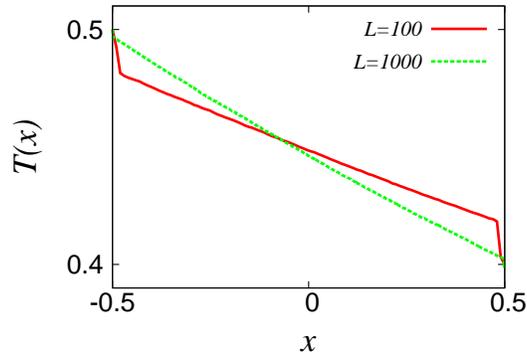}
\end{picture}
\caption{Temperature profiles at $L=100$ and $1000$ in the scaled space $x=\frac{i}{L}-\frac{1}{2}$.
$(T_1, T_2)=(0.5, 0.4)$, which corresponds to $(\Tbar,\varepsilon)=(0.45,0.222)$.
For small lattices such as $L=100$, greater gaps between the temperature at  both boundaries exist, and the slope in the bulk is lower. The gaps become smaller, and the slope 
becomes almost $(T_1-T_2)/L$ for larger lattices, as exemplified by $L=1000$.}
\label{fig:Tprofile}
\end{figure}

In  Figs.~\ref{fig:SleqSneq} and \ref{fig:extensivity}, we have also plotted the thermodynamic limit of $\Sleq(L,\Tbar,\varepsilon)$, which will be derived in the next section
as \eqref{e:SleqInf}.
We find that both $\Sneq(L,\Tbar,\varepsilon)$ and $\Sleq(L,\Tbar,\varepsilon)$
deviates from the thermodynamic limit.
As shown in Fig.~\ref{fig:Tprofile},
there exist gaps in the temperature profile at the boundaries of the lattice,
which are a typical finite-size effect,  becoming smaller as $L$ increases
and vanishing in the thermodynamic limit.
Their coincidence is barely disturbed by the finite-size effect
or the imposed temperature gradient.
This holds, even for a lattice with $L=10$, in which the temperature gradient is never small.
Thus, we expect that $\Di\Sneq$ converges to the dotted line as $L\rightarrow\infty$,
keeping the coincidence $\Di\Sneq\simeq\Di\Sleq$.
As already mentioned, the good coincidence holds even at $\varepsilon\simeq 0.55$,
where the temperature profile is not as linear as the line in Fig.~\ref{fig:Tprofile},
and the values of $\Di\Sneq$ and $\Di\Sleq$ greatly deviate from the thermodynamic limit, as shown in Fig.~\ref{fig:SleqSneq}.
Note that the local quantities for small systems are difficult to access in real experiments,
whereas they are measurable in numerical experiments.
Because the entropy productions in \eqref{e:S2} are a global quantity
measured just at the boundaries of the system,
our method is applicable to determine the local quantities for such small systems,
and moreover, 
because the SST possess wider applicability than the local equilibrium,
our method offers new experimental possibilities for complex systems whose
local quantities are difficult to define.

\section{Extensivity and additivity for the nonequilibrium entropy}

\begin{figure}
\centering
\begin{picture}(370,250)
\includegraphics[scale=0.3]{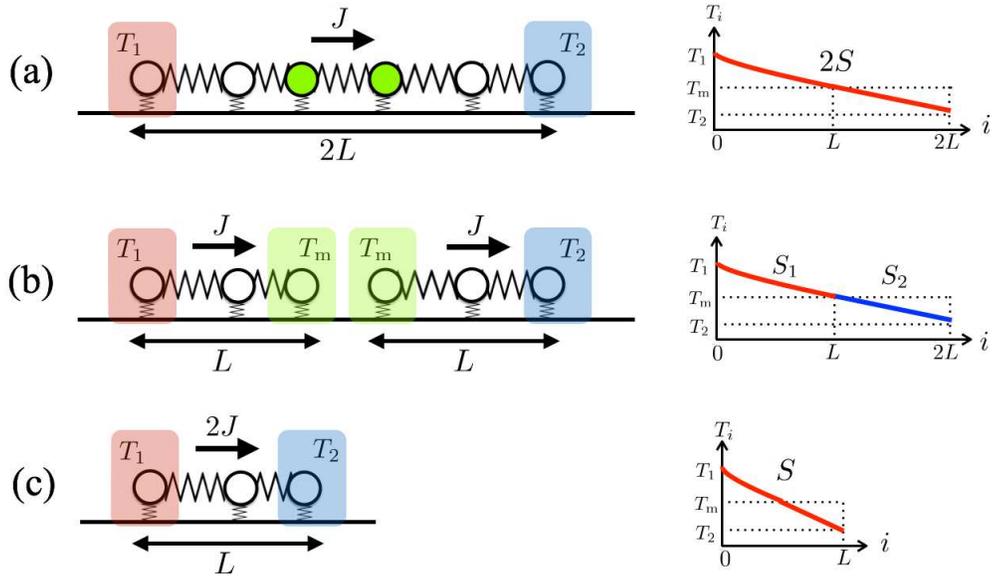}
%\begin{picture}(220,150)
%\includegraphics[scale=0.27]{ExtAdd.eps}
\end{picture}
\caption{Schematic illustrating the difference between the extensivity and the additivity in the heat-conducting states. In the separation of  the system in (a) into (b), we first attach heat baths at an appropriate temperature of $T_\mathrm{m}$ at $i=L$ and $L+1$, conserving the value of the heat current $J$. Then, we remove the interaction between the two. We have the additivity of the entropy as $2S=S_1+S_2$; however, $S_1\neq S_2 \neq S$. We have the extensivity of the system in (a) from (c), where $J$ is not conserved.}
\label{fig:ExtAdd}
\end{figure}

In this section, we discuss the system in the thermodynamic limit $L\rightarrow \infty$
under steady heat conduction.
As is well known, 
the extensivity and additivity are  the most important properties in equilibrium thermodynamics.
Noting their importance,
Sasa and Tasaki proposed their SST in \cite{Sasa-Tasaki}
from an argument of the extensivity and additivity in the NESSs.
They examined the additivity of the heat-conducting systems 
and proposed intensive and extensive variables for the degree of nonequilibrium.
The steady heat current $J$ was considered as a candidate for a new intensive variable,
and its conjugate extensive variable was defined by ${\partial F}/{\partial J}$
via an extended thermodynamic function $F$.
Their SST seem to be powerful as an extension of thermodynamics to the NESSs,
containing the second law or the convexity of the thermodynamic functions
and predictions of the NESSs.

The entropy defined from the excess entropy production has been considered to be the basic quantity for the SST in \cite{Oono-Paniconi, Hatano-Sasa, Sasa-Tasaki, KNST}.
Therefore, we examine the validity of the  assumptions by Sasa and Tasaki 
on SST from the viewpoint of $\Sneq$ determined from the observed excess entropy production.
First, we examine the additivity of the heat-conducting system and then proceed to the extensivity.
Naively speaking, the additivity is expected to hold
because we identified the coincidence of $\Sneq$ with $\Sleq$.

In the thermodynamic limit, we can apply the equation of heat conduction.
The $\phi_4$ model is known to be a normal heat conducting medium that satisfies Fourier's law,
$J(x) =-\lambda(T(x))\nabla T$, where $J(x)$ and $\lambda(T)$ are the heat current at a position $x$ and the heat conductivity as a function of $T$, respectively.
We took a scaled space $x=\frac{i}{L}-\frac{1}{2}$ with the boundary conditions 
$T(x=-\frac{1}{2})=\TL$ and $T(x=\frac{1}{2})=\TR$.
Our numerical simulation suggests that $\lambda(T)\simeq\lambda_0 {T}^{-4/3}$
with a constant $\lambda_0$, which is consistent with the scaling reported in \cite{Aoki-Kusnezov-conductivity}.
In the steady state, $J(x)=\Jst$ for any position $x$
as a steady solution of the heat-conduction equation $\nabla J(x)=0$.
We obtain the steady temperature profile as
\begin{align}
T(x) =
\left[ 1-\varepsilon x +\frac{2\varepsilon^2}{3}\left(x^2-\frac{1}{4}\right)\right]\Tbar 
+O(\varepsilon^3)
.
\label{e:Tprofile}
\end{align}
The steady heat current in the original space indexed by $i$ becomes
\begin{align}
\Jst(L,\Tbar, \varepsilon)=\lambda_0\Tbar^{-\frac{1}{3}} \frac{\varepsilon }{L}+O(\varepsilon^3)
,
\label{e:Jst}
\end{align}
where $L$ is the size of the lattice.
The numerically obtained temperature profile agrees with the estimated $T(x)$ in \eqref{e:Tprofile} for sufficiently long lattices, 
as is exemplified by the line of $L=1000$ in Fig.~\ref{fig:Tprofile},
whereas the profile begins to exhibit gaps in the temperature for small lattices, 
as is shown in the line of $L=100$ in the same figure.
These gaps are known to be due to the finite-size effect in the $\phi_4$ model.

We demonstrate the additivity by separating a lattice with a size of $2L$ into two lattices with a size of $L$, as shown in Fig. \ref{fig:ExtAdd}.
In order to maintain the steady heat current $\Jst$ before and after the separation,
we attach new heat baths at a temperature of $T_\mathrm{m}$ at the ends of the two shorter lattices 
at $i=L/2$ (Fig. \ref{fig:ExtAdd}(b)).
Taking $T_\mathrm{m}:=T(x=0)=(1-\frac{\varepsilon^2}{6})\Tbar+O(\varepsilon^3)$,
as assigned by the steady solution in \eqref{e:Tprofile} of the heat-conduction equation,
we have 
\begin{align}
\Jst(2L)=\Jst_\mathrm{left}(L)=\Jst_\mathrm{right}(L)
\end{align}
with an error of $O(\varepsilon^3)$.
Thus, we can separate the lattice into two without changing its local properties or temperature profile,
and $\Jst$ can be an intensive variable, as was proposed by Sasa and Tasaki.

As a next step, we examine the additivity of the entropy.
Integrating the local entropy $s(x)=c \log T(x)$ over the space as $\Sleq=L \int^{1/2}_{-1/2} s(x) dx$,
 we have the following estimate for the thermodynamic limit:
\begin{eqnarray}
\Sneq(L,\Tbar,\varepsilon) =\Seq(L,\Tbar)-\frac{11}{72}c L \varepsilon^2+O(\varepsilon^3)
,
\label{e:SleqInf}
\end{eqnarray}
where we applied $\Sneq(L,\Tbar,\varepsilon)=\Sleq(L,\Tbar,\varepsilon)$.
We now adopt a new notation $S(L,T_1,T_2;J):=\Sneq(L,\Tbar,\varepsilon)$ to emphasize the boundary conditions
to drive the system to the NESSs and steady heat current.
Let us compare the entropies for the original and separated lattices. 
From \eqref{e:SleqInf},  we have
\begin{align}
S(2L,T_1,T_2;J)&=\Seq(2L,\Tbar)-\frac{11}{36}cL\varepsilon^2+O(\varepsilon^3),\\
S(L,T_1,T_\mathrm{m};J)&=\Seq\left(L, \left(1+\frac{\varepsilon}{4}-\frac{\varepsilon^2}{12}\right)\Tbar\right)-\frac{11}{72}cL\frac{\varepsilon^2}{4}+O(\varepsilon^3),\\
S(L,T_\mathrm{m},T_2;J)&=\Seq\left(L, \left(1-\frac{\varepsilon}{4}-\frac{\varepsilon^2}{12}\right)\Tbar\right)-\frac{11}{72}cL\frac{\varepsilon^2}{4}+O(\varepsilon^3).
\end{align}
It is remarkable that $S(L,T_1,T_\mathrm{m};J)\neq S(L,T_\mathrm{m},T_2;J)$ owing to the differrence in $\Seq$.
Substituting $\Seq(L,T)=cL\log T$ and applying the Taylor expansion, 
we obtain the additivity such that
\begin{align}
S(2L, T_1, T_2; J)=S(L, T_1, T_\mathrm{m}; J)+S(L, T_\mathrm{m},T_2; J)
\label{e:add}
\end{align}
with a precision of $O(\varepsilon^2)$.

\begin{figure}
\centering
\begin{picture}(350,100)
\includegraphics[scale=0.26]{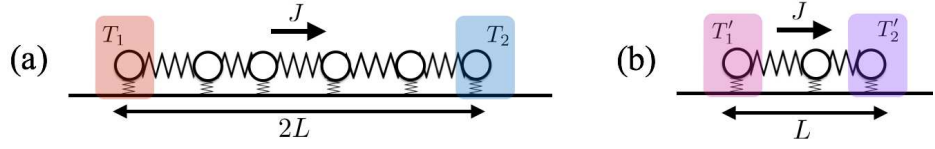}
%\begin{picture}(220,150)
%\includegraphics[scale=0.27]{ExtAdd.eps}
\end{picture}
\caption{The examination of the extensivity when conserving the heat current $J$.
The systems in (a) and (b) are assumed to have the same mean temperature $\Tbar$ and heat current $J$.
$T_1=(1 +\frac{\varepsilon}{2})\Tbar$ and $T_2=(1 -\frac{\varepsilon}{2})\Tbar$,
whereas 
$T_1'=(1+\frac{\varepsilon}{4})\Tbar $ and $T_2'=(1 -\frac{\varepsilon}{4})\Tbar $.
We have $S(2L)>2S(L)$,  which means the failure of the extensivity for the conditions of constant $J$ and $\Tbar$.}
\label{fig:Ext2}
\end{figure}

Next, we examine the extensivity. 
The estimate in \eqref{e:SleqInf} indicates that $\Sneq$ is extensive,
i.e., $\Sneq(L,\Tbar,\varepsilon)\propto  L$.
We compare this extensivity with the one assumed by Sasa and Tasaki \cite{Sasa-Tasaki} 
such that
\begin{align}
S_\mathrm{ST}( n L,T, J)=nS_\mathrm{ST}(L,T, J)
,
\label{e:extST}
\end{align}
where $S_\mathrm{ST}$ is the entropy with thermodynamic variables $(L,T,J)$
following  the argument by Sasa and Tasaki.
Although
the extensivity formulated in \eqref{e:SleqInf} holds for systems at fixed temperatures $T_1$ and $T_2$,
$J$ cannot be constant for fixed $T_1$ and $T_2$
but is inversely proportional to $L$, i.e.,
\begin{align}
S(nL, T_1, T_2;  J/n)&= n S(L, T_1, T_2; J),
\label{e:ext}
\end{align}
as is schematically drawn in Figs. \ref{fig:ExtAdd}(a) and (c).
The scaling in \eqref{e:ext} is distinct from that in \eqref{e:extST}.

To clarify the inconsistency between the scaling in \eqref{e:ext} and that in \eqref{e:extST},
 let us consider two lattices (see Fig. \ref{fig:Ext2}): 
One has a size of $2L$ attached to baths with temperatures of $T_1=(1+\frac{\varepsilon}{2})\Tbar$ and $T_2=(1-\frac{\varepsilon}{2})\Tbar $,
and the other has a size of $L$ attached to baths with temperatures of $T_1'=(1+\frac{\varepsilon}{4})\Tbar $ and $T_2'=(1-\frac{\varepsilon}{4})\Tbar $.
The two lattices have the same mean temperature $\Tbar$ and steady heat current $J$,
whereas the size differes by a factor of two, as is proposed in the scaling in \eqref{e:extST}.
According to the estimate in \eqref{e:SleqInf}, we have
\begin{align}
S(2L, T_1, T_2; J)=\Sneq(2L,\Tbar,\varepsilon)>2 S(L, T_1', T_2'; J) = 2 S(L,\Tbar,\frac{\varepsilon}{2})
\end{align}
which is not the same as the assumed scaling in \eqref{e:extST}.
Thus, we conclude that we do not have the extensivity proposed in \cite{Sasa-Tasaki}.
Remembering that the heat conduction in the $\phi_4$ model is as normal as satisfying Fourier's law, the scaling in \eqref{e:extST} would not hold in other realistic systems,
regardless of the functional form of the heat conductivity. 
The scaling in \eqref{e:extST} does not hold, even if the conductivity is a constant independent of the temperature.

From the examination above, we realize the complexity of the extensivity and additivity in the NESSs. 
This is in contrast to the equilibrium states, where
the additivity is degenerate into the extensivity.
The entropy for the heat-conducting states is characterized by its extensivity and additivity as well as the equilibrium states.
However,
the extensivity does not degenerate into the additivity.
This is a significant difference between the equilibrium and nonequilibrium entropies.

\section{Discussion}

We performed numerical experiments to determine  $\Sneq$
for the heat-conducting states in one-dimensional lattices exhibiting Fourier's law.
We discovered that $\Sneq$ connected to the excess heat currents 
coincides with $\Sleq$ determined according to the local equilibrium hypothesis.
This coincidence indicates that the SST based on the excess heat
are categorized as an inclusive description for the local equilibrium hypothesis.
The major points of the former may be the wider applicability than the local equilibrium regardless of its complexity or size and  the accessibility of the observable.
Note that  $\Theta$ and $\dTheta$ in \eqref{e:S2} are accessible in any system.
What one should measure is the heat currents at the boundaries of the system,
not the local quantities inside of the system,
along the forward and backward protocols.
Our finding suggests that 
thermodynamic quantities consistent with the local equilibrium
become experimentally studied,
 even for systems whose local states are not defined, not accessible, 
or greatly deviated from the thermodynamic limit.

In equilibrium, 
the extensivity and additivity are degenerated and not distinguished.
We confirmed that the two properties were extended to the heat-conducting states,
but we simultaneously found a breakdown of the degeneration between the two properties.
This raises the question of whether the conventional logic in thermodynamics can be applied 
to nonequilibirum thermodynamics.
For instance, Sasa and Tasaki \cite{Sasa-Tasaki} assumed
the extensivity and additivity for the NESSs
 to apply several powerful methods used in equilibriurm thermodynamics.
Since they do not seem to recognize the  breakdown of the degeneration,
their approach and predictions should be reconsidered.
On the other hand, the formulae for the additivity in \eqref{e:add}  and extensivity in \eqref{e:ext} are similar to 
those formulated by Bodineau and Derrida \cite{Bodineau-Derrida}, 
which were introduced in the large deviation functional for the current fluctuations.
Their functional may be related to  $\Sneq$, even though the initial approach is different,.
It may be possible to unify these two approaches, i.e., the large deviation functionals and the operational relation.

Acknowledgement:
We are grateful to Keiji Saito for useful comments and suggestions.
N.N. also thanks T.S. Komatsu, S-i. Sasa and H. Tasaki for various related discussions 
on SST.  This work was supported by KAKENHI Nos. 15K05196 and 25103002.

\section{Appendix}

\subsection{Derivation of \eqref{e:S2}}

The relation in \eqref{e:S2} has already been derived in \cite{KNST};
however, we derive it from the Jarzynski-type equality in \eqref{e:KNST} for the NESSs, 
which was derived as eq.~(3.1) by Komatsu et al. in \cite{KNST-exact}.
We restrict our derivation to  protocols that operate at the temperatures $\Tb_1$ and $\Tb_2$.
Letting $\beta_1(t):=1/\Tb_1(t)$, $\beta_2(t):=1/\Tb_2(t)$,
the mean inverse temperature $\beta:=(\beta_1+\beta_2)/2$ be the reference inverse temperature of  sysrem,
and $\varepsilon(t)=(\beta_2(t)-\beta_1(t))/\beta(t)=(\Tb_1(t)-\Tb_2(t))/\Tbar(t)$,
we introduce the nonequilibrium part of the entropy production rate and
the entropy production due to the heat current as
\begin{align}
\psi(t;\hat\Gamma) &:= 
\varepsilon(t) j_{\psi}(t;\hat\Gamma)
,\\
j_{\psi}(t;\hat\Gamma)&:=\beta(t)\frac{j_1(t;\hat\Gamma)-j_2(t;\hat\Gamma)}{2},
\end{align}
where $j_1(t;\hat\Gamma)$ and $j_2(t;\hat\Gamma)$ are heat currents from the respective heat baths to the system at time $t$
in a trajectory $\hat\Gamma$.
We also introduce
\begin{align}
\phi(t;\hat\Gamma) :=\psi(t;\hat\Gamma)+\dot\beta H(\Gamma(t))
,
\end{align}
with which we have
\begin{align}
\Phi[\hat\Gamma]&=\Theta[\hat\Gamma]+\beta(\tau)H(\Gamma(\tau))-\beta(0)H(\Gamma(0))
,
\label{e:PT}
\end{align}
where 
the accumulation of $\phi$ over the period of the protocol, say $0\le t \le \tau$,  is denoted as
\begin{align}
\Phi[\hat\Gamma] = \int_0^{\tau} \phi(t;\hat\Gamma) ~dt
.
\end{align}
In the protocol shown in  Fig.~\ref{fig:protocol}, the temperature deviates at most of $O(\varepsilon)$ so that we can write $\beta(t)=\beta(0)+\varepsilon b(t)$, 
which leads to $\dot\beta=\varepsilon \dot b$.
 Thus, we have an estimate $\Phi[\hat\Gamma]=O(\varepsilon)$.
For a later purpose, we define $X[\hat\Gamma]$ such that
$\Phi[\hat\Gamma]=\varepsilon X[\hat\Gamma]$, which corresponds to
\begin{align}
X[\hat\Gamma]:= 
\int_0^{\tau} 
\left\{
\frac{\varepsilon(t)}{\varepsilon} j_{\psi}(t;\hat\Gamma) 
+\dot b(t) H(\Gamma(t))
\right\}~dt
.
\label{e:X}
\end{align}

In the following, we write the set of operational parameters as $\alpha$,
and $\alpha=(\Tb_1,\Tb_2)$ in the present Paper.
For a precise description, let us denote
$(\alpha)$, $\hat\alpha$, and $\hat\alpha^\dagger$ 
as the constant,  forward and  backward protocols for changing the degree of nonequilibiurm $\varepsilon$, 
where $\hat\alpha$ is the protocol for changing from $\alpha$ to $\alpha'$.
Then, the extended Jarzynski equality derived in \cite{KNST-exact} is written as
\begin{align}
M(\alpha')-M(\alpha)=\log\frac{\bbkt{e^{-\frac{\Phi}{2}}}^{\hat\alpha}}{\bbkt{e^{-\frac{\Phi}{2}}}^{\hat\alpha^\dagger}},
\label{e:KNST}
\end{align}
where $\sbkt{\cdot}^{\hat\alpha}$ and $\sbkt{\cdot}^{\hat\alpha^\dagger}$ are the path averages along the respective protocols.
$M(\alpha)=-\beta F(\alpha)$, where $F(\alpha)$ is the nonequilibrium free energy adopted in \cite{KNST-exact},
 which corresponds to the normalization factor for the steady probability density in the NESSs.
In the protocol in Fig.~\ref{fig:protocol}, $\alpha=(\Tbar,0)$ and $\alpha'=(\Tbar,\varepsilon)$.

Letting the fluctuation in $\Psi$ from the mean be
\begin{align}
\delta\Phi[\hat\Gamma]
:=\Phi[\hat\Gamma]-\bbkt{\Phi}^{\hat\alpha}
,
\end{align}
the relation in \eqref{e:KNST} is written as
\begin{align}
M(\alpha')-M(\alpha)+\frac{1}{2}\left(\bbkt{\Phi}^{\hat\alpha}-\bbkt{\Phi}^{\hat\alpha^\dagger}\right)
=\log\bbkt{e^{-\frac{\delta\Phi}{2}}}^{\hat\alpha}-\log\bbkt{e^{-\frac{\delta\Phi}{2}}}^{\hat\alpha^\dagger}
,
\end{align}
which corresponds to
\begin{align}
S(\alpha')-S(\alpha)+\frac{1}{2}\left(\bbkt{\Theta}^{\hat\alpha}-\bbkt{\Theta}^{\hat\alpha^\dagger}\right)
=\log\bbkt{e^{-\frac{\delta\Phi}{2}}}^{\hat\alpha}-\log\bbkt{e^{-\frac{\delta\Phi}{2}}}^{\hat\alpha^\dagger}
\label{e:KNST2}
\end{align}
by substituting \eqref{e:PT} and 
$M(\alpha)=S(\alpha)-\beta U(\alpha)$.
Applying the cumulant expansion to the r.h.s. of \eqref{e:KNST2}, we have
\begin{align}
\log\bbkt{e^{-\frac{\delta\Phi}{2}}}^{\hat\alpha}-\log\bbkt{e^{-\frac{\delta\Phi}{2}}}^{\hat\alpha^\dagger}
=\frac{1}{8}\left(\bbkt{\Phi;\Phi}^{\hat\alpha}-\bbkt{\Phi;\Phi}^{\hat\alpha^\dagger}\right)+O(\varepsilon^3)
,
\end{align}
where $\sbkt{A;B}=\sbkt{AB}-\sbkt{A}\sbkt{B}$.

Since the order of the change in the protocol in Fig.~\ref{fig:protocol} is of $O(\varepsilon)$,
we have $\sbkt{A}^{\hat\alpha}=\sbkt{A}^{(\alpha)}+O(\varepsilon)$ for any observable $A(\hat\Gamma)$.
Therefore, we approximate the two culumlants as
\begin{align}
&\bbkt{\Phi;\Phi}^{\hat\alpha}=\bbkt{\Phi;\Phi}^{(\alpha)}+O(\varepsilon^3),\\
&\bbkt{\Phi;\Phi}^{\hat\alpha^\dagger}=\bbkt{\Phi;\Phi}^{(\alpha)}+O(\varepsilon^3),
\end{align}
which indicate that the order of the r.h.s. of \eqref{e:KNST2} is of $O(\varepsilon^3)$.
Thus, we reach the relation in \eqref{e:S2}.

\subsection{Comment on \eqref{e:delta}}

Next, we explain the degree of the deviation from the quasistatic limit $\delta$
in \eqref{e:delta}.
For general protocols with a finite speed of operation, we have
\begin{align}
\bbkt{\Thetae}^{\hat\alpha}+\bbkt{\Thetae}^{\hat\alpha^\dagger}
&=
\bbkt{\Phie}^{\hat\alpha}+\bbkt{\Phie}^{\hat\alpha^\dagger}\nonumber\\
&=\varepsilon (\bbkt{\Xe}^{\hat\alpha}+\bbkt{\Xe}^{\hat\alpha^\dagger})
,
\label{e:tr}
\end{align}
which is obtained by substituting 
$\sbkt{\Phie}^{\hat\alpha}=\sbkt{\Thetae}^{\hat\alpha}+\beta(\tau)U(\alpha')-\beta(0) U(\alpha)$ resulting from \eqref{e:PT}
and then \eqref{e:X}.
Therefore, we can write
\begin{align}
\bbkt{\Thetae}^{\hat\alpha}+\bbkt{\Thetae}^{\hat\alpha^\dagger}=\varepsilon\delta,\\
\delta :=\bbkt{\Xe}^{\hat\alpha}+\bbkt{\Xe}^{\hat\alpha^\dagger},
\end{align}
where $\delta$ is a quantity that depends on the protocol.
Noting that the mean of any excess quantity vanishes in the steady states $\sbkt{\Xe}^{(\alpha)}=0$, we have $\delta=0$ in the quasistatic limit.
Thus, we recognize that $\delta$ indicates the degree of the deviation from the quasistatic limit.

Remembering that $\sbkt{\Xe}^{\hat\alpha}=\sbkt{\Xe}^{(\alpha)}+O(\varepsilon)$
in general protocols for changing $\varepsilon$ and $\sbkt{\Xe}^{(\alpha)}=0$,
we have
\begin{align}
\bbkt{\Xe}^{\hat\alpha}
=\bbkt{\Xe}^{(\alpha)}+O(\varepsilon)
=O(\varepsilon)
,
\end{align}
which corresponds to 
\begin{align}
\delta=O(\varepsilon)
.
\end{align}
This means that the relation in \eqref{e:S3} connected entropy to the excess entropy production
has worse precision than \eqref{e:S2} in general.


\begin{thebibliography}{10}

%% SST 

\bibitem{Landauer}
R. Landauer,
 %{\em $dQ= TdS$ far from equilibrium}\/, 
Phys. Rev. A18, 255-266 (1978).


\bibitem{Oono-Paniconi}
{Y. Oono and M. Paniconi},
{Prog. Theor. Phys. Suppl.}
\textbf{130}, 29 (1998).

\bibitem{Ruelle}
D. Ruelle,
%{\em Extending the definition of entropy to nonequilibrium steady states}\/,
Proc. Natl. Acad. Sci. U.S.A.  {\bf 100}, 3054--3058 (2003).\\
%{\tt arXiv:cond-mat/0303156}


\bibitem{Sasa-Tasaki}
S.-i. Sasa and H. Tasaki,
%{\em Steady State Thermodynamics}\/,
J. Stat. Phys. {\bf 125}, 125--224 (2006).\\
%{\tt arXiv:cond-mat/0411052}




%Thermodynamic relations in NESS
\bibitem{Hatano-Sasa}
{T. Hatano and S.-i. Sasa,}
%{\em Steady-State Thermodynamics of Langevin Systems}\/,
%{\tt arXiv:cond-mat/0010405}
{Phys. Rev. Lett.}
\textbf{86}, 3463 
{(2001)}.



%\bibitem{BGJLL2012}
%L. Bertini, D. Gabrielli, G. Jona-Lasinio, and C. Landim,
%{\em Thermodynamic transformations of nonequilibrium states}\/,
%J. Stat. Phys. {\bf 149}, 773--802 (2012).\\
%{\tt arXiv:1206.2412}



%% KNST 

\bibitem{KNST}
T. S. Komatsu, N. Nakagawa, S.-i. Sasa and H. Tasaki,
%{\em Steady State Thermodynamics for Heat Conduction --- Microscopic Derivation}\/,
Phys. Rev. Lett. {\bf 100}, 230602 (2008).\\ 
%{\tt arXiv:0711.0246}

\bibitem{KNST-nl}
T. S. Komatsu, N. Nakagawa, S.-i. Sasa and H. Tasaki,
%{\em Entropy and Nonlinear Nonequilibrium Thermodynamic Relation for Heat Conducting Steady States}\/,
J. Stat. Phys. {\bf 142}, 127--153 (2011).\\ %{\tt arXiv:1009.0970}

\bibitem{Saito-Tasaki}
K. Saito and H. Tasaki,
%{\em Extended Clausius Relation and Entropy for Nonequilibrium Steady States in Heat Conducting Quantum Systems}\/,
J. Stat. Phys. {\bf 145}, 1275--1290 (2011).\\
%{\tt arXiv:1105.2168}

\bibitem{NN2012}
N. Nakagawa,
%{\em Work Relation and the Second Law of Thermodynamics in Nonequilibrium Steady States}\/,
Phys. Rev. E {\bf 85}, 051115 (2012).\\
%{\tt arXiv:1109.1374}

\bibitem{KNST-exact}
T. S. Komatsu, N. Nakagawa, S. -i. Sasa and H. Tasaki,
%{\em Exact Equalities and Thermodynamic Relations for Nonequilibrium Steady States}
J. Stat. Phys. {\bf 158}, 1195-1414 (2015).



\bibitem{BGJLL2013}
L. Bertini, D. Gabrielli, G. Jona-Lasinio, and C. Landim,
%{\em Clausius Inequality and Optimality of Quasistatic Transformations for Nonequilibrium Stationary States}\/,
Phys. Rev. Lett. {\bf 110}, 020601 (2013).\\
%{\tt arXiv:1208.1872}


\bibitem{MaesNetocny2014}
C. Maes and K. Netocny,
%{\em A nonequilibrium extension of the Clausius heat theorem}\/,
J. Stat. Phys. {\bf 154}, 188--203 (2014).\\
%{\tt arXiv:1206.3423}

\bibitem{Spinney-Ford}
R. E. Spinney and I. J. Ford, 
Phys. Rev. Lett. {\bf 108}, 170603 (2012).\\

\bibitem{Spinney-Ford2}
R. E. Spinney and I. J. Ford, 
Phys. Rev. E  {\bf 85}, 051113 (2012).

\bibitem{Sasa2013}
S.-I. Sasa,
%{\em Possible extended forms of thermodynamic entropy}\/,
J. Stat. Mech.  P01004 (2014).\\
%{\tt arXiv:1309.7131}

%%%%%%%

%phi4 model
\bibitem{Aoki-Kusnezov}
K. Aoki and D. Kusnezov, 
Annals of Physics {\bf 295}, 50–80 (2002)
%Physics Letters B {\bf 477}, 348 (2000).

%Stochastic energetics

\bibitem{Sekimoto-JPSJ}
{K. Sekimoto},
{J. Phys. Soc. Jpn.}
\textbf{66},
{1234}
{(1997)}.

\bibitem{Sekimoto}
K. Sekimoto, {\em Stochastic Energetics}
{(Springer, Berlin, 2010)}.



\bibitem{Aoki-Kusnezov-conductivity}
K. Aoki and D. Kusnezov, Phys. Lett. A {\bf 265}, 250 (2000).

\bibitem{KNST-numerical}
T. S. Komatsu, N. Nakagawa, S. Sasa, H. Tasaki and N. Ito,
{Prog. Theor. Phys. Suppl.}
\textbf{184},
{329} {(2010)}


%\bibitem{Lebowitz-Spohn}
%J. L. Lebowitz and H. Spohn,
%{\em A Gallavotti-Cohen-Type Symmetry in the Large Deviation Functional for Stochastic Dynamics}
%J. Stat. Phys. {\bf 95}, 333-365 (1999).




\bibitem{Bodineau-Derrida}
T. Bodineau and B. Derrida,
%{\em Current Fluctuations in Nonequilibrium Diffusive Systems: An Additivity Principle}\/,
Phys. Rev. Lett. {\bf 92}, 180601 (2004).\\

\bibitem{Aoki-Kusnezov-LEQ}
K. Aoki and D. Kusnezov, Phys. Lett. A \textbf{309}, 377 (2003).



%Geometric effect in excess heat

%\bibitem{SagawaHayakwa}
%T. Sagawa and H. Hayakawa,
%{\em Geometrical expression of excess entropy production}\/,
%Phys. Rev. E {\bf 84}, 051110 (2011).\\
%{\tt arXiv:1109.0796}

%\bibitem{YugeSagawaSugitaHayakawa}
%T. Yuge, T. Sagawa, A. Sugita, and H. Hayakawa,
%{\em Geometrical Excess Entropy Production in Nonequilibrium Quantum Systems}, J. Stat. Phys. {\bf 153}, 412--441 (2013).\\
%{\tt arXiv:1305.5026}

%\bibitem{NN2014}
%N. Nakagawa,
%{\em Universal exact expression for adiabatic pumping in terms of nonequilibrium steady states}\/,
%Phys. Rev. E {\bf 60}, 022108 (2014).\\
%{\tt arXiv:1401.4242}





%%  KN
%\bibitem{KN}
%{T. S. Komatsu and N. Nakagawa},
%Phys. Rev. Lett. {\bf 100}, 030601 (2008);

%\bibitem{KN-long}
%T. S. Komatsu, N. Nakagawa, S. Sasa and H. Tasaki,
%J. Stat. Phys. {\bf 134}, 401 (2010).



\end{thebibliography}
\end{document}